\documentclass[aps,prl,          groupedaddress,showpacs,twocolumn,amsmath,amssymb]{revtex4}
\usepackage[dvips]{graphicx}
\newcommand{\VEC}[1]{\overrightarrow{#1}}
\newcommand{\pdd}[2]{\displaystyle \frac{\partial #1}{\partial #2}}
\newcommand{\ds}{\displaystyle}
\begin{document}


\title{Apparent violation of equipartition of energy in constrained dynamical systems}


\author{Tetsuro Konishi}
\affiliation{Department of Physics, Nagoya University, Nagoya, 464-8602, Japan}

\author{Tatsuo Yanagita}
\affiliation{Research Institute  for Electronic Science, Hokkaido University, Sapporo,
001-0020, Japan}


\date{\today}

\begin{abstract}

We propose
a planar chain system, which is a simple mechanical system  with a constraint. 
It is composed of $N$ masses connected by $N-1$ light links.
%
It can be considered as a model of a  chain system, e.g., a polymer,
in which each bond is replaced by  a rigid link.
The long time average of the kinetic energies of the masses in this model is numerically computed.
It is found that the average kinetic energies of the masses are different and masses near the ends
of the chain have large energies.
We explain that this result is not in contradiction with the principle of equipartition.
The apparent violation of equipartition is observed not only in the planar chain systems but also in other constrained systems.
We derive an approximate expression for the average kinetic energy,
which is in qualitative agreement with the numerical results. 
\end{abstract}


\maketitle

Information on energy distribution in many-body systems is quite important 
 for both  theoretical and practical purposes.
If a system is in thermal equilibrium, then, 
according to the principle of equipartition of energy, 
the average kinetic energy is equally distributed among all the degrees of freedom.
Even when this principle holds, however, we found that nonuniform distribution of the average 
kinetic energy can occur.

In this letter, we introduce a system called a ``planar chain system'',
which is a simplified model of a chain system e.g., a polymer.
We show that the average kinetic energy in this system is 
nonuniformly distributed even when it is in thermal equilibrium, 
but the  principle of equipartition is not violated. 
We explain  the reason 
for the  nonuniform distribution of energy, 
which we refer to as the ``{\it apparent violation of 
equipartition of energy}''.
This property of apparent violation of equipartition of energy could provide a   new insight into
the behavior of chain systems. 




 \begin{figure}
 \includegraphics[width=6truecm]{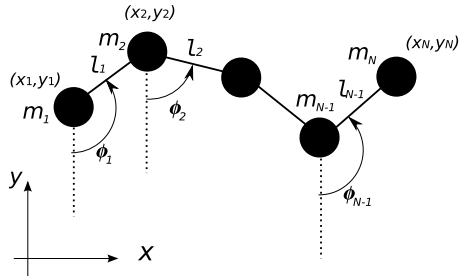}
 \caption{\label{fig:planar-chain}A planar chain system}
\end{figure}
Let us introduce the planar chain system. 
The planar chain system is composed of  $N$ particles (masses) connected by  $N-1$ links.
The masses can rotate smoothly, as shown in Fig.\ref{fig:planar-chain}. 
The links are massless and have fixed lengths.
The system  is defined by the following Lagrangian $L$ and constraints $g_i$ ($i=1,2,\cdots, N-1$): 
\begin{align}
  L&=\sum_{i=1}^N \frac{m_i}{2}\left(\dot x_i^2 + \dot y_i^2\right)
- U(\left\{ \VEC{r_i} \right\} ) \ , \label{eq:chain-Lag} \\
g_i(\left\{\VEC{r_j}\right\}) &\equiv \frac{1}{2}
\left\{ \left|\VEC{r_{i+1}}-\VEC{r_i}\right|^2 - \ell_i^2\right\}  = 0,  
\label{eq:chain-constraint}
\end{align}
where $N$ is the number of particles, 
$m_i$ is the mass of $i$'th particle, $\VEC{r_i}\equiv(x_i,y_i)$ represents the
position of  the $i$'th particle, and $\ell_i$ is the length of the $i$'th link. 
$U$ represents potential energy.
We consider (i) a free chain with $U\equiv 0$ and
(ii)  external potential  $U\equiv \sum_{i=1}^N V(\VEC{r_i})$.

If we define $\varphi_i$ as the angle between the $i$'th link and the $-y$
direction (Fig.\ref{fig:planar-chain}), we can rewrite the Lagrangian
without the  constraint.
First we consider the following relations:
$    x_{i+1} - x_i = \ell_i \sin\varphi_i$ , \ 
$    y_{i+1} - y_i = -\ell_i \cos\varphi_i \ . $
  Using the total mass $M$ and the center of mass $(X_G, Y_G)$ defined as
$ M\equiv \sum_{i=1}^N m_i,$
$ X_G\equiv\sum_{i=1}^N  \frac{m_i}{M} x_i,$ \ 
$ Y_G\equiv \sum_{i=1}^N \frac{m_i}{M} y_i, $
we obtain
\begin{equation}
  x_i = X_G + \sum_{j=1}^{N-1}a_{ij}\sin\varphi_j  \, , \ \ 
  y_i = Y_G - \sum_{j=1}^{N-1}a_{ij}\cos\varphi_j  \, , 
\end{equation}
where  $a_{ij}$ is defined as
\begin{equation}
  a_{ij}\equiv
  \begin{cases}
\displaystyle
\phantom{-}\mu_j^{\le}\,\ell_j
&
 \ : \  j < i \ , \\ 
%
\displaystyle
- \mu_j^{>}\,\ell_j
&
\ :\  j \ge i  \ , \\ 
  \end{cases}
  \label{eq:chain-a-ij}
\end{equation}
and 
\begin{equation}
\mu_k\equiv \frac{m_k}{M}, \   \  \mu_n^{\le} \equiv \sum_{k=1}^n \mu_k \ ,  \  \ 
 \mu_n^{>} \equiv \sum_{k=n+1}^N \mu_k \ .
\label{eq:mu-gl}
\end{equation}
By a  straightforward calculation, we obtain 
the Lagrangian (\ref{eq:chain-Lag})
in terms of  $\varphi_i$'s  and  $(X_G,Y_G)$ as
\begin{align}
L
&=\frac{M}{2}(\dot X_G^2 + \dot Y_G^2)
+
\frac{M}{2}\!\sum_{j,k=1}^{N-1} 
A_{jk}(\varphi)
\dot\varphi_j\dot\varphi_k  \nonumber\\
& - U(X_G, Y_G, \left\{\varphi_i\right\})\
\ , 
\label{eq:chain-ke-varphi-2} \\
A_{jk}(\varphi)&\equiv\mu_{\min(j,k)}^{\le}\,\mu_{\max(j,k)}^{>}\,\cos(\varphi_{jk})\ell_j \ell_k \ , 
\label{eq:M-and-mu}
\end{align}
where $\varphi_{jk}\equiv \varphi_j - \varphi_k$.

We can consider this system as a simplified prototype of various chain
systems, e.g., proteins, polymers and  spacecraft manipulators,
under the assumption that the frequencies of bond-stretching vibrations are quite high.




Now, we describe a method for numerical simulation.
The Lagrangian that is expressed in terms of  angles (\ref{eq:chain-ke-varphi-2}) is complicated and 
it is difficult to numerically integrate the equation of motion, 
in particular for large $N$. 
Hence, we use the original form of the Lagrangian 
(\ref{eq:chain-Lag}) and the constraint $g_i$ (\ref{eq:chain-constraint}).
Then, the equation of motion includes  terms of the  
constraint, which is called a ``Lagrange multiplier''~\cite{goldstein}. 
We determine Lagrange multipliers
numerically at each integration step  so that the  constraint is satisfied~\cite{leimkurler-reich}.
Methods of this type, e.g.,  ``SHAKE'' and ``RATTLE'' algorithms, 
are  widely used for molecular simulation in chemistry
~\cite{constrained-md-1,constrained-md-2,CHARMM}.
In addition, some of the algorithms are known to be symplectic~\cite{leimkurler-reich}.
Here, we use the forth-order symplectic integrator. 
In  some cases, we verify  the results by using an implicit Runge-Kutta method.

If $U\equiv 0$, the total angular momentum 
is conserved, hence, in this case,
the energy distribution is different from the 
microcanonical distribution. 
In actual computations, we place the system in
a potential wall composed of arcs of radius $a$:
$U\equiv \sum_{i=1}^NV(\VEC{r_i})$, $V(\VEC{r})=0.01 \sum_{j=1}^{N_\text{wall}}
\left|\left|\VEC{r} - \VEC{R}_\text{j}\right|-a\right|^{-6}$. 
Then,  the system 
exhibits strongly chaotic motion similar to  billiards~\cite{billiards-book} and does not have
any conserved quantities other than the total energy, hence the microcanonical distribution is restored.
\begin{figure}[htbp]
  \centering
  \includegraphics[width=4cm]{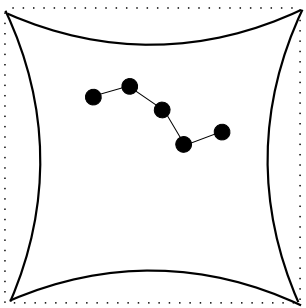}
  \caption{Planar chain in potential wall. 
   $N=5$, $N_\text{wall}=4$,
  $a=4N\ell$,
    $R_j=(R,0),(-R,0),(0,R),(0,-R)$, $R=N\ell + \sqrt{a^2-N^2\ell^2}$.
    $\ell_1=\cdots=\ell_{N-1}\equiv \ell$.
  }
  \label{fig:chain-in-billiards}
\end{figure}

\begin{figure}[htbp]
  \centering
  \includegraphics[width=5cm]{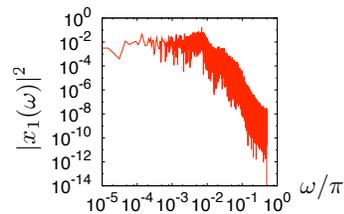}
  \caption{Power spectrum of $x_1(t)$ for  $0\le t \le 32767$. 
    $N=5$. External potential in Fig.\protect\ref{fig:chain-in-billiards} is used
    to obtain the spectrum.}
  \label{fig:power-x1}
\end{figure}

Although the planar chain system is a simple system, 
its dynamics is complex; further, energy exchanges occur between various 
parts of the system.
Fig.~\ref{fig:power-x1} shows a power spectrum of $x_1(t)$ with the external potential
mentioned above.
It is a  broad continuous spectrum, which is a manifestation
of chaotic motion~\cite{Lichtenberg-Lieberman}.

Using the method described above, we compute the long time average of 
kinetic energy.
If the averaging time is sufficiently large, the  long time average and thermal average
can be assumed to be  the same.

The kinetic energy of $i$'th  particle is defined as
$  K_i(t)
\equiv
\frac{m_i}{2}\left(\dot x_i^2 + \dot y_i^2\right) \, ,
$
and its long time average is defined as
\begin{equation}
  \label{eq:ke-linear-lt-ave}
  \overline{K_i} \equiv
  \frac{1}{t_\text{max}}\int_0^{t_\text{max}}K_i(t)\,dt
 \ , \ t_{\text{max}}\rightarrow \infty \ . 
\end{equation}

\begin{figure}[htbp]
  \centering
  \includegraphics[width=7truecm]{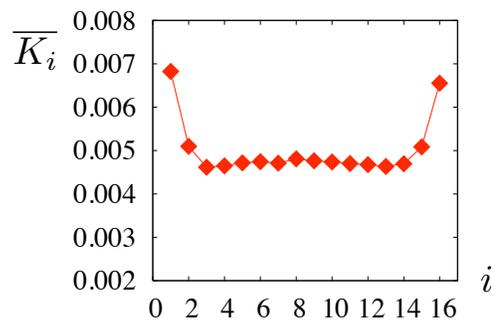}
  \caption{Long time average of  kinetic energy  $\overline{K_i}$ vs. $i$
  (\ref{eq:ke-linear-lt-ave}). $N=16$. 
  $m_i = 1.0$ for all $i$\ and  \ $\ell_i = 1.0$ for all $i$.
  The initial condition is as follows: $x_i=(i-1)-N/2$, $y_i=0$,
$p^{(x)}_{i}=0$ for all $i$, $p^{(y)}_{1}=-0.1$,  and $p^{(y)}_{i}=0.1 ( i>1)$.
Here, $p^{(x)}_{i}$ and $p^{(y)}_{i}$ represent the $x$ and $y$ components of 
the momentum of the $i$'th particle, respectively.
%
The time step for integration is $dt=0.001$.
 $t_\text{max}$ (\ref{eq:ke-linear-lt-ave}) is $10^5$.
The relative error for total energy (square root of  the time average of the squared displacement) is 
$\sqrt{\overline{\Delta E^2/E_0^2}}=6.1\times 10^{-11}$.
}
  \label{fig:ke-linear-lt-ave}
\end{figure}

\begin{figure}[htbp]
  \centering
  \includegraphics[width=6truecm]{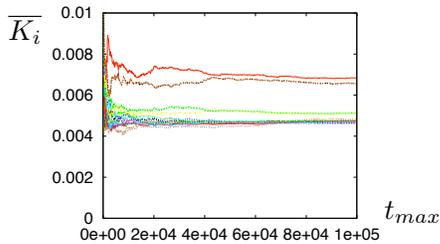}
  \caption{Convergence of  kinetic energy  $K_i$
  (\ref{eq:ke-linear-lt-ave})  in Fig.\ref{fig:ke-linear-lt-ave} as a function
  of $t_\text{max}$.
}
  \label{fig:ke-linear-lt-convergence}
\end{figure}

One might think that the values of all  $\overline{K_i}$'s in this system 
must be the same, by regarding $K_i$ as the kinetic
energy of the $i$-th degree of freedom and applying the principle of
equipartition of  energy.
However, this is not true.
Fig.\ref{fig:ke-linear-lt-ave} shows a plot of 
the average kinetic energy of each mass $\overline{K_i}$ (\ref{eq:ke-linear-lt-ave}) against $i$
for $N=16$ planar chain system. 
It is clear that the $\overline{K_i}$'s are not equally distributed.
More importantly, we find that masses that are  near the ends of the chain  have 
large kinetic energies. 
We obtain this result for  all the computed  system sizes  ( $N\le 64$ ).
Fig.~\ref{fig:ke-linear-lt-convergence} shows the  convergence of $
\frac{1}{t_\text{max}}\int_0^{t_\text{max}}K_i(t')dt'$ as a function of $t_\text{max}$. 
The values shown in 
Fig.\ref{fig:ke-linear-lt-ave} are well converged.

However, this remarkable result is not in contradiction with the
principle of equipartition of energy.
The principle of equipartition of energy 
is stated as follows~\cite{Kubo-book}:
Suppose we have a system
defined by a Hamiltonian
\begin{equation}
     H(q,p) \equiv K(q,p) + V(q) , \ 
     K(q,p) \equiv \sum_{i,j=1}^{\mathcal{N}}\frac{1}{2}\alpha_{ij}(q)p_i p_j \ \  , 
 \end{equation}
where $p_i$ and $q_i$ are canonically conjugate to each other
and $\mathcal{N}$ is  the total number of degrees of freedom. 
If  it is in  thermal equilibrium at temperature $T$, 
then  the following relation holds:
\begin{equation}
\left<
\frac{1}{2}p_i \pdd{K}{p_i}
\right>
= \frac{1}{2}k_B T
\label{eq:equipartition}
\end{equation}
(Summation over the index $i$ is not taken in the left hand side.).
The symbol $\left< \cdots \right>$ represents thermal average at
$T$, and is defined as
\begin{equation}
  \left<
f(q,p)
\right>
\equiv
\frac{1}{Z}\int f(q,p) e^{-\beta H}d\Gamma \ ,
 \label{eq:thermal-ave-def}
\end{equation}
for any function $f(q,p)$.
Here, $d\Gamma$ is a volume element of phase space, $Z$ is a partition
function, and $\beta\equiv 1/k_B T$.

Let us define the ``canonical kinetic energy'' $K_i^{(c)}$ and
the ``linear kinetic energy'' $K_i$ as 
\begin{equation}
 K_i^{(c)} \equiv \frac{1}{2}p_i \pdd{K}{p_i}, \ \ \ \ 
K_i\equiv\frac{1}{2}m_iv_i^2,
\label{eq:ke-canonical-and-linear-def}  
\end{equation}
respectively.
Here, equipartition of energy means that the average values of  $K_i^{(c)}$'s 
 are equal at thermal equilibrium.

For systems such as  gas models 
or lattice models, 
$\alpha_{ij}(q)=m_i^{-1}\delta_{ij}$
and $K_i^{(c)}= K_i$: hence,
the principle of equipartition~(\ref{eq:equipartition})
simply means that
$
  \left<\frac{1}{2}m_i v_i^2\right> = \frac{1}{2}k_B T \ ,
$
which is a commonly used form of equipartition of energy.

However, in the case of a  planar chain system,
equipartition of energy has a different meaning.
From ~(\ref{eq:chain-ke-varphi-2}),
we obtain the canonical momentum $p_i$ that is conjugate to  $\varphi_i$ as
$  p_i\equiv\frac{\partial L}{\partial \dot \varphi_i}
=\sum_{k=1}^{N-1} A_{ik}(\varphi)\dot\varphi_k \ , 
$
and we  obtain the canonical kinetic energy $K_i^{(c)}$  for the
planar chain system as
\begin{equation}
K_i^{(c)}\equiv \frac{1}{2}p_i \pdd{K}{p_i} = \frac{1}{2}
\sum_{k=1}^{N-1} A_{ik}(\varphi)\dot\varphi_k \dot\varphi_i\ .
\label{eq:ke-canonical}
\end{equation}
For example, eq. (\ref{eq:ke-canonical}) with $N=3$ and $i=1$ we have 
\begin{equation}
  K_1^{(c)}
= \frac{M}{2}\left\{
\mu_1(\mu_2+\mu_3)\ell_1^2\dot\varphi_1^2
+
\mu_1\mu_3\ell_1\ell_2\dot\varphi_1\dot\varphi_2
\cos\varphi_{12}
\right\} \ .
\end{equation}

It should be noted that $K_i^{(c)}$ is defined by variables of every part of the 
system, whereas $K_i$ is defined only by the $i$-th particle.
In other words, canonical kinetic energy $K_i^{(c)}$  is extended,
whereas linear kinetic energy $K_i$ is localized.
Hence it is obvious that $K_i^{(c)} \ne K_i$. Since  $K_i^{(c)}$
obeys equipartition of energy, we can consider that $K_i$
does not obey this principle.

\begin{figure}[htbp]
  \centering
  \includegraphics[width=7truecm]{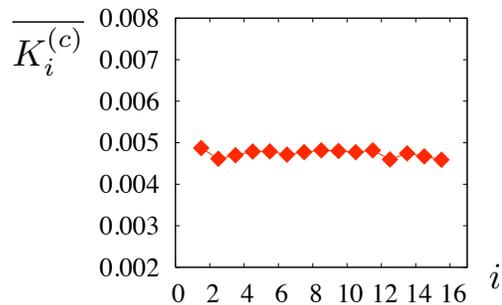}
  \caption{Long time average of canonical kinetic energy
  $\overline{K_i^{(c)}}$ (\ref{eq:ke-canonical}) vs $i$.
  This plot is obtained by using the same data as that in  Fig.\ref{fig:ke-linear-lt-ave}.
}
  \label{fig:ke-canonical-lt-ave}
\end{figure}

Fig.~\ref{fig:ke-canonical-lt-ave}  shows the long time average of 
$K_i^{c}$ (\ref{eq:ke-canonical}) 
for the same time series as that in Fig.~\ref{fig:ke-linear-lt-ave}.
It is clearly shown that 
$\overline{K_i^{(c)}}$\,'s take almost the same value for all $i$. 
That is, equipartition of energy is realized.
Non-equipartition of energy clearly shown in 
Fig.\ref{fig:ke-linear-lt-ave} does not imply 
that principle of equipartition is violated.
In other words, although the system obeys the principle of  equipartition of energy,
the values of the linear kinetic energy are different at different points in the system.
We refer to the variation in  $\left<K_i\right>$  under thermal equilibrium  as the
``{\it apparent violation of equipartition of energy}''.


We derive the nonuniformity of the energy distribution by
analytical calculation.
By a straightforward calculation, we obtain
\begin{align}
\frac{\left< K_i \right>}{k_B T}
&
=
\frac{m_i}{M}
+
\frac{m_i}{2k_B T}
\sum_{j,k=1}^{N-1}
a_{ij}a_{ik}
\left<
\cos\left(\varphi_{jk}\right)
\dot\varphi_j\dot\varphi_k
\right>
  \ .
\label{eq:chain-ke-g-i}
\end{align}


To evaluate the second term, we adopt the following  approximations:
\begin{align}
\left<
\cos\left(\varphi_{jk}\right)
\dot\varphi_j\dot\varphi_k
\right> = 0 \ \ \text{for} \ j\ne k \ , 
\label{eq:chain-diag-approx} \\
\left(A^{-1}\right)_{jj}\sim \frac{1}{A_{jj}} = \frac{1}{M\mu_j^{\le}\mu_j^{>}\ell_j^2}  \ \ .
\end{align}
(The matrix $A^{-1}$ is included in $\exp(-\beta H)$.)
These approximations indicate that each link in the chain
rotates independently.
Then, we obtain
\begin{equation}
\frac{  \left< K_i \right>}{k_B T}
=
\frac{m_i}{M}
\left\{
1 + 
\frac{1}{2}
\left[
\sum_{j=1}^{i-1} 
\left(\frac{\mu_j^{\le}}{ \mu_j^{>}} 
\right)
+\sum_{j=i}^{N-1}
\left(\frac{\mu_j^{>}}{\mu_j^{\le}} 
\right)
\right]
\right\} \, .
\label{eq:chain-ke-d-approx-2}
\end{equation}
Details of the calculation will be shown elsewhere~\cite{konishi-yanagita-chain-2}.

Eq.(\ref{eq:chain-ke-d-approx-2}) shows that  the average
linear kinetic energy $ \left<K_i\right>$
varies from point to point. In other words, 
``apparent  violation  of equipartition of  energy'' occurs in this system.


If all the masses are the same
$
m_i = m, 
$
then  we obtain
\begin{equation}
\frac{  \left<
K_i
\right>}
{k_B T}
=
\frac{1}{N}
\left\{
1 + 
\frac{1}{2}
\left[
\sum_{j=1}^{i-1} 
\left(\frac{j}{N-j} 
\right)
+\sum_{j=i}^{N-1}
\left(\frac{N-j}{j} 
\right)
\right]
\right\} \, .
\label{eq:chain-ke-ave-equalmass}  
\end{equation}
This expression implies the following:
\begin{equation}
  \langle K_1\rangle
> \langle K_2\rangle
> \cdots
< \langle K_{N-1}\rangle
< \langle K_N\rangle \ .
\end{equation}
It is clear that $\langle K_i\rangle$ is
large at  the ends  of the chain and small at the center of the the chain:
this result is in  qualitative agreement with  the result of the 
 numerical computation shown in Fig.\ref{fig:ke-linear-lt-ave}.


In this letter we have numerically shown that 
$\ds K_i\equiv \frac{m_i}{2}v_i^2$ does not obey the principle of
equipartition of energy
for the planar chain system.
 Moreover $\overline{K_i}$ of particles that are near both ends  of the chain is large.
The nonuniform distribution of the linear kinetic energy is qualitatively
explained by analytical calculation.

The apparent contradiction is due to  the difference between 
 $K_i\equiv\frac{1}{2}m_iv_i^2$ and
 $K_i^{c}\equiv \frac{1}{2}p_i\pdd{K}{p_i}$.
This difference is caused by the presence 
of the coordinate $q$ ($\varphi$ for planar chain systems) 
in the expression of the  kinetic energy $K$,  
due to the existence  of the constraint.
Further,  the same numerical time series show that the average values of 
$K_i^{(c)}$ 
are equal, i.e., the system obeys the principle of equipartition.



It is clear that there are other models in which the  values of 
the average kinetic energy are not equal.
These models are systems with constraints, where the expression of the kinetic energy includes coordinates.
In fact, it has been found that the behavior of linear kinetic energy in 
a multiple pendulum system is similar to that in the planar chain system ~\cite{yanagita-gakkai-1};
and we will report detailed analysis elsewhere~\cite{konishi-yanagita-pendulum}.
In polymer science, the three-dimensional version of this model is known as
a ``freely jointed chain''~\cite{Kramers-chain,Mazers-chain-pre-1996}.
We expect that 
the behavior of the kinetic energy in the freely jointed chain will be similar to 
that in the planar chain system.

We have shown that in the  planar chain system, the energy at the ends of the
chain is larger than that at the center. This result may be considered rather trivial,
because it may seem that the end parts can be moved easily.
However,
even in thermal equilibrium,  where all degrees of freedom have
the same energy on average, the energy at the ends of the chain is large.

This result would
have important implications for the dynamics of chain systems 
such as molecules, proteins, polymers, and some artificial objects.
For example, in polymer science, it is well known  that atoms situated near the ends of 
the polymer chain have characteristic behavior called the ``end effect''~\cite{dotera}.
Apparent violation of equipartition of energy we found in planar chain systems can be closely related to the origin of the end effect of polymers.

\begin{acknowledgments}
T.K. would like to thank M. Toda, Y. Y. Yamaguchi, T. Komatsuzaki, T. Dotera and K. Nozaki for fruitful discussions.
This study was partially supported by a Grant-in-Aid for Scientific Research (C)
(20540371) from the Japan Society for the Promotion of Science (JSPS).
\end{acknowledgments}

\end{document}